\documentclass[journal]{IEEEtran}
\usepackage{graphicx}
\title{Geometrical theory of whispering gallery modes}
\author{Michael L. Gorodetsky, Aleksey E. Fomin\thanks{M.V. Lomonosov Moscow State University, Moscow, Russia}}

\begin{document}
\maketitle

\begin{abstract}
Using quasiclassical approach rather precise analytical
approximations for the eigenfrequencies of whispering gallery modes
in convex axisymmetric bodies may be found. We use the eikonal
method to analyze the limits of precision of quasiclassical
approximation using as a practical example spheroidal dielectric
cavity. The series obtained for the calculation of eigenfrequencies
is compared with the known series for dielectric sphere and with
numerical calculations. We show how geometrical interpretation
allows expansion of the method on arbitrary shaped axisymmetric
bodies.
\end{abstract}


\section{Introduction}

Submillimeter size optical microspheres made of fused silica with whispering
gallery modes (WGM)\cite{microspheres} can have extremely high quality-factor,
up to $10^{10}$ that makes them promising devices for applications in
optoelectronics and experimental physics. Historically Richtmyer\cite{Richtmyer} was the first to suggest
that whispering gallery modes in axisymmetric dielectric body should have very
high quality-factor. He examined the cases of sphere and torus.
However only recent breakthroughs in technology in
several labs allowed producing not only spherical and not only fused silica but
spheroidal, toroidal \cite{microtorus,Vahala} or even arbitrary form
axisymmetrical optical microcavities from crystalline materials preserving or
even increasing high quality factor \cite{crystall}.
Especially interesting are devices manufactured of nonlinear optical crystals.
Microresonators of this type can be used as high-finesse cavities
for laser stabilization, as frequency discriminators and high-sensitive
displacement sensors, as sensors  of ambient medium and in optoelectronical
RF high-stable oscillators. (See for example materials of the special LEOS workshop on
WGM microresonators \cite{LEOS}).

The theory of  WGMs in microspheres is well established and allows precise
calculation of eigenmodes, radiative losses and field distribution both
analytically and numerically. Unfortunately, the situation changes drastically
even in the case of simplest axisymmetric geometry, different from ideal
sphere or cylinder. No closed analytical solution can be found in this case.
Direct numerical methods like finite elements method are also inefficient when
the size of a cavity is several orders larger than the wavelength. The theory
of quasiclassical methods of eigenfrequencies approximation starting from
pioneering paper \cite{Keller} have made a great progress lately \cite{Babich}.
For the practical evaluation of precision that these methods can in principal
provide, we chose a practical problem of calculation of  eigenfrequencies in
dielectric spheroid and found a series over angular mode number $l$.
This choice of geometry is convenient due to several reasons:
1) other shapes, for example toroids \cite{Vahala} may be approximated by
equivalent spheroids; 2) the eikonal equation as well as scalar Helmholtz equation
(but not the vector one!) is separable in spheroidal coordinates that gives additional
flexibility in understanding quasiclassical methods and comparing them with
other approximations; 3) in the limit of zero eccentricity spheroid turns to
sphere for which exact solution and series over $l$ up to $l^{-8/3}$ is known \cite{Schiller}.

The Helmholtz vector equation is unseparable \cite{TETM} in spheroidal
coordinates and no vector harmonics tangential to the surface of spheroid
can be build. That is why there are no pure TE or TM modes in spheroids but
only hybrid ones. Different methods of separation of variables (SVM) using series
expansions with either spheroidal or spherical functions have been proposed
\cite{Asano,Farafonov,Charalambopoulos}. Unfortunately they lead
to extremely bulky infinite sets of equations which can be solved numerically
only in simplest cases and the convergence is not proved.
Exact characteristic equation for the eigenfrequencies in
dielectric spheroid was suggested\cite{Moraes} without provement that if real
could significantly ease the task of finding eigenfrequencies. However, we can
not confirm this claim as this characteristic equation contradicts limiting cases
with the known solutions i.e. ideal sphere and axisymmetrical oscillations in
a spheroid with perfectly conducting walls \cite{Closed}.

Nevertheless, in case of whispering gallery modes adjacent to equatorial plane
the energy is mostly concentrated in tangential or normal to the surface
electric components that can be treated as quasi-TE or quasi-TM modes and analyzed
with good approximation using scalar wave equations.

\section{Spheroidal coordinate system}

There are several equivalent ways to introduce prolate and oblate
spheroidal coordinates and corresponding eigenfunctions
\cite{Komarov,Li,Abramowitz}. The following widely used system of coordinates
allows to analyze prolate and oblate geometries simultaneously:
\begin{eqnarray}
&&x=\frac{d}{2}[(\xi^2-s)(1-\eta^2)]^{1/2}\cos(\phi)
\nonumber\\
&&y=\frac{d}{2}[(\xi^2-s)(1-\eta^2)]^{1/2}\sin(\phi)\nonumber\\
&&z=\frac{d}{2}\xi\eta,
\end{eqnarray}
where we have introduced a sign variable $s$ whish is equal to 1 for the prolate geometry with
$\xi\in[1,\infty)$ determining spheroids and $\eta\in[-1,1]$ describing
two-sheeted hyperboloids of revolution (Fig.1, right). Consequently, $s=-1$ gives
oblate spheroids for $\xi\in[0,\infty)$ and one-sheeted hyperboloids of
revolution  (Fig.1, right). $d/2$ is the semidistance between focal points. We are
interested in the modes inside a spheroid adjacent to its surface in
the equatorial plane. It is convenient to designate a semiaxis in this
plane as $a$ and in the $z$-axis of rotational symmetry of the body as
$b$. In this case $d^2/4=s(b^2-a^2)$ and eccentricity
$\varepsilon=\sqrt{1-(a/b)^{2s}}$. The scale factors for this system
are the following:
\begin{eqnarray}
h_\xi&=&\frac{d}{2}\left(\frac{\xi^2-s\eta^2}{\xi^2-s}\right)^{1/2},\nonumber\\ h_\eta&=&\frac{d}{2}\left(\frac{\xi^2-s\eta^2}{1-\eta^2}\right)^{1/2},\\ h_\phi&=&\frac{d}{2}[(\xi^2-s)(1-\eta^2)]^{1/2}.\nonumber
\end{eqnarray}

\begin{figure*}
\centering
\includegraphics[height=7cm]{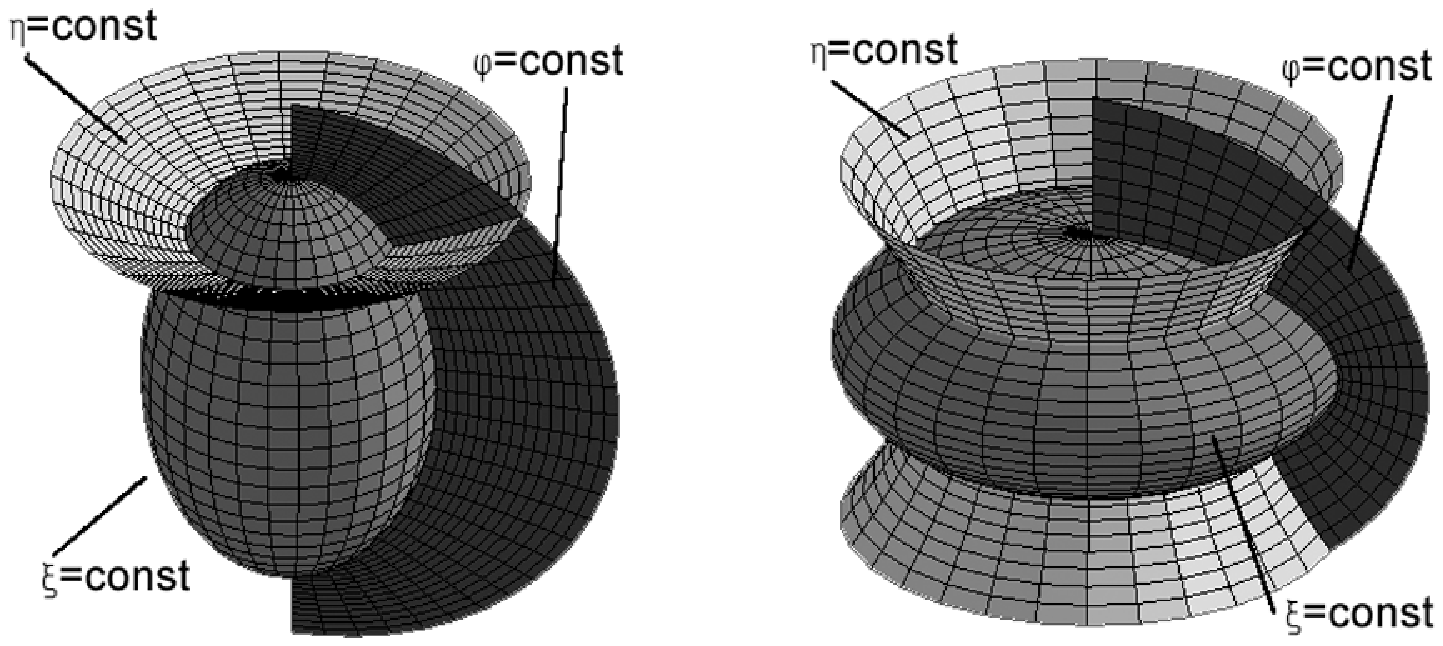}
\caption{Graphical representation of prolate and oblate spheroidal coordinate systems ($\xi,~\eta,~\phi$).}
\end{figure*}

The scalar Helmholtz differential equation  is separable
\begin{equation}
\Delta \Phi + k^2 \Phi=0.
\end{equation}
\begin{eqnarray}
\frac{\partial}{\partial \xi}
(\xi^2-s)\frac{\partial}{\partial \xi}\Phi+\frac{\partial}{\partial
\eta}(1-\eta^2)\frac{\partial}{\partial \eta}\Phi \nonumber\\
+\left(c^2(\xi^2\mp\eta^2)-\frac{m^2}{1-\eta^2}-s\frac{m^2}{\xi^2-s}\right)
\Phi =0,
\end{eqnarray}
where $c=kd/2$. The solution is
$\Phi=R_{ml}(c,\xi)S_{ml}(c,\eta)e^{im\phi}$ where radial and angular
functions are determined by the following equations:
\begin{eqnarray}
\frac{\partial}{\partial \xi} (\xi^2-s)\frac{\partial R}{\partial
\xi}-\left(\lambda_{ml}-c^2\xi^2+s\frac{m^2}{\xi^2-s}\right) R =0 \label{radial},
\end{eqnarray}
\begin{eqnarray}
\frac{\partial}{\partial \eta}(1-\eta^2)\frac{\partial S}{\partial
\eta} +\left(\lambda_{ml}-sc^2\eta^2-\frac{m^2}{1-\eta^2}\right) S =0. \label{angular}
\end{eqnarray}
Here $\lambda_{ml}$ is the separation constant of the
equations which should be independently determined and it is a function
on $m$, $l$ and $c$. With substitution $\xi=2r/d$ the first equation
transforms to the equation for the spherical Bessel function $j_l(kr)$
if $d/2\to 0$ in which case the second equation immediately turns to
the equation for the associated Legendre polynomials $P_m^l(\eta)$ with
$\lambda=l(l+1)$. That is why spheroidal functions are frequently
analyzed as decomposition over these spherical functions.

The calculation of spheroidal functions and of $\lambda_{ml}$ is not a trivial
task \cite{MoraesR,MoraesS}. The approximation of spheroidal functions and their zeroes may seem more
straightforward for the calculation of eigenfrequencies of spheroids, however
we found that another approach that we develop below gives better results and
may be easily generalized to other geometries.

\section{Eikonal approximation in spheroid}

The eikonal approximation is a powerful method for solving optical
problems in inhomogeneous media where the scale of the variations is
much larger than the wavelength. It was shown by Keller and
Rubinow \cite{Keller} that it can also be applied to eigenfrequency problems
and that it has very clear quasiclassical ray interpretation.
It is important that this quasiclassical ray interpretation requiring
simple calculation of the ray paths along the geodesic surfaces and application
of phase equality (quantum) conditions gives precisely the same equations
as the eikonal equations. Eikonal equations allow, however, to obtain more
easily not only eigenfrequencies but field distribution also.

In the eikonal approximation the solution of the Helmholtz scalar
equation is found as superposition of straight rays:
\begin{eqnarray}
u({\bf r})=A({\bf r})e^{\imath k_0 S({\bf r})}.
\end{eqnarray}
The first order approximation for the phase function $S$
called eikonal is determined by the following equation.
\begin{eqnarray}
(\nabla S)^2=\epsilon({\bf r}),
\end{eqnarray}
where $\epsilon$ is optical susceptibility. For our problem of
searching for eigenfrequencies $\epsilon$ does not depend on coordinates,
$\epsilon=n^2$ inside the cavity and $\epsilon=1$ -- outside.
Though the eikonal can be found as complex rays in the external area
and stitched on the boundary as well  as ray method of Keller and
Rubinow \cite{Keller,Babich,GDT} can be extended for whispering  gallery
modes in dielectrical bodies in a more rigorous way \cite{Silakov}.
To do so we must account for an additional phase shift on the dielectric
boundary. Fresnel amplitude coefficient of reflection \cite{Born}:
\begin{eqnarray}
{\cal R}=\frac{\chi\cos\theta-i\sqrt{n^2\sin^2\theta-1}}{\chi\cos\theta+i\sqrt{n^2\sin^2\theta-1}},
\end{eqnarray}
where $\chi=1$ for quasi-TE modes and $\chi=1/n^2$ for quasi-TM modes.
give the following approximations for the phase shift for grazing angles:
\setlength{\arraycolsep}{0.0em}
\begin{eqnarray}
i \ln{\cal R}&=&\pi-\Theta_r\\
&\simeq&\pi-\frac{2\chi}{\sqrt{n^2-1}}\cos\theta-\frac{\chi(3n^2-2\chi^2)}{3(n^2-1)^{3/2}}\cos^3\theta\nonumber\\
&-&\frac{\chi(15n^4-20n^2\chi^2+8\chi^4)}{20(n^2- 1)^{5/2}}\cos^5\theta + O(\cos^7\theta) \nonumber \label{Goos}
\end{eqnarray}
\setlength{\arraycolsep}{5pt}
However direct use of this phase shift in the equations for internal rays
as suggestes in \cite{Silakov} leads to incorrect results. The reason is a
well known Goos-H\"anchen effect -- the shift of the reflected beam along the
surface. The beams behave as if they are reflected from the surface hold away
from the real boundary at $\sigma_r=\frac{\Theta_r}{2k\cos\theta}$.
That is why we may substitute the problem for a dielectric body with the
problem for an equivalent body enlarged on $\sigma_r$ with the totally
reflecting boundaries. The parameters of equivalent spheroid are marked below
with overbars.

The eikonal equation separates in spheroidal coordinates  if we choose
$S=S_1(\xi)+S_2(\eta)+S_3(\phi)$:
\begin{eqnarray}
\frac{\xi^2-s}{\xi^2-s\eta^2}\left(\frac{\partial S_1(\xi)}{\partial
\xi}\right)^2+\frac{1-\eta^2}{\xi^2-s\eta^2}\left(\frac{\partial S_2(\eta)
}{\partial \eta}\right)^2+\nonumber\\
+\frac{1}{(\xi^2-s)(1-\eta^2)}\left(\frac{\partial S_3(\phi)
}{\partial \phi}\right)^2=\frac{n^2d^2}{4}.
\end{eqnarray}
After immediate separation of $\left(\frac{\partial S}{\partial
\phi}\right)=\mu$ we have:
\begin{eqnarray}
(\xi^2-s)\left(\frac{\partial S_1(\xi)}{\partial
\xi}\right)^2+(1-\eta^2)\left(\frac{\partial S_2(\eta) }{\partial
\eta}\right)^2+\nonumber\\
+\frac{s\mu^2}{\xi^2-s}+\frac{\mu^2}{1-\eta^2}
-\frac{n^2d^2}{4}(\xi^2-s\eta^2)=0.
\end{eqnarray}
Introducing another separation constant $\nu$ we finally
obtain solutions:
\begin{eqnarray}
&&\frac{\partial S_1(\xi)}{\partial
\xi}=\left(\frac{n^2d^2\xi^2}{4(\xi^2-s)}-\frac{\nu^2}{\xi^2-s}-\frac{s\mu^2}{(\xi^2-s)^2}\right)^{1/2},\nonumber\\
&&\frac{\partial S_2(\eta)}{\partial\eta}
=\left(\frac{\nu^2}{1-\eta^2}-\frac{n^2d^2s\eta^2}{4(1-\eta^2)}-\frac{\mu^2}{(1-\eta^2)^2}\right)^{1/2},
\end{eqnarray}
which after some manipulations transform to:
\begin{eqnarray}
&&\frac{\partial S_1(\xi)}{\partial\xi}
=\frac{nd}{2}\frac{\sqrt{(\xi^2-\xi_r^2)(\xi^2-s\eta_r^2)}}{\xi^2-s}\nonumber\\
&&\frac{\partial S_2(\eta)}{\partial\eta}
=\frac{nd}{2}\frac{\sqrt{(\eta_r^2-\eta^2)(\xi_r^2-s\eta^2)}}{1-\eta^2}\nonumber\\
&&\frac{\partial S_3(\phi)}{\partial\phi}=\mu \label{eikonal},
\end{eqnarray}
where
\begin{eqnarray}
\eta_r^2&=&\frac{(1+s\alpha)-\sqrt{(1+s\alpha)^2-4s\alpha\eta^2_0}}{2s\alpha}\nonumber\\
\xi_r^2&=&\frac{(1+s\alpha)+\sqrt{(1+s\alpha)^2-4s\alpha\eta^2_0}}{2\alpha}
\nonumber\\
&=&\frac{1+s\alpha}{\alpha}-s\eta^2_r,
\end{eqnarray}
where $\alpha=\frac{n^2d^2}{4\nu^2}$, $\eta^2_0=1-\mu^2/\nu^2$.
It is now the time to turn to the quasiclassical ray interpretation \cite{Keller,Babich}.
The equation for the eikonal describes the rays that can spread inside spheroid along the
straight line. These are the rays that freely go inside spheroid than touch
the surface and reflect. For the whispering gallery modes the angle of reflection
is close to $\pi/2$. The closest to the center points of these rays form the caustic
surface which is the ellipsoid determined by a parameter $\xi_r$. The rays are the tangents to this
internal ellipsoid and follow along the geodesic lines on it. In case of ideal sphere
all the rays of the same family lie in the same plane. However, even a slightest
eccentricity removes this degeneracy and inclined closed circular modes which
should be more accurately called quasimodes \cite{Arnold} are turned into
open-ended helices winding up on caustic spheroid precessing \cite{precess}, and
filling up the whole region as in a clew. The upper and lower points of these
trajectories determine other caustic surface with a parameter $\eta_r$ determining
two-sheeted hyperbolid for prolate or one-sheeted hyperboloid for oblate spheroid. The value of
$\eta_r$ has very simple mechanical interpretation. The rays in the eikonal
approximation are equivalent to the trajectories of a point-like billiard ball inside the
cavity. As axisymmetrical surface can not change the angular momentum related to the $z$
axis, it should be conserved as well as the kinetic energy (velocity).
That is why $\eta_r$ is simply equal to the sine of the angle between the equatorial
plane and the trajectory crossing the equator and at the same time it determines
the maximum elongation of the trajectory from the equator plane. If all the rays touch the
caustic or boundary surface with phases that form stationary distribution
(that means that the phase difference along any closed curve on them
is equal to integer times $2\pi$), then the eigenfunction and hence
eigenfrequency is found.

To find the circular integrals of phases $kS$ (\ref{eikonal}) we should take into account
the properties of phase evolutions on caustic and reflective boundary. Every touching
of caustic adds $\pi/2$ (see for example \cite{Babich}) and reflection adds $\pi$. Thus for
$S_1$ we have one caustic shift of $\pi/2$ at $\xi_r$ and one reflection from the
equivalent boundary surface $\xi_s$ (at the distance $\sigma$ from the real surface),
for $S_2$ -- two times $\pi/2$ due to caustic shifts at $\pm\eta_r$, and
we should add nothing for $S_3$.
\begin{eqnarray}
&&k\Delta S_1=2k\int\limits_{-\xi_r}^{\xi_s}\frac{\partial S_1}{\partial\xi}d\xi=2\pi(q-1/4)\nonumber\\
&&k\Delta S_2=2k\int\limits_{-\eta_r}^{\eta_r}\frac{\partial S_2}{\partial\eta}d\eta=2\pi(p+1/2)\nonumber\\
&&k\Delta S_3=k\int\limits_0^{2\pi}\frac{\partial S_3}{\partial\phi}d\phi=2\pi |m|,\label{integrals}
\end{eqnarray}
where $q=1,2,3 ...$ -- is the order of the mode, showing the number of the zero of
the radial function on the surface, and $p=l-|m|=0,1,2...$.
These conditions plus integrals (\ref{eikonal}) completely coincide with those obtained by Bykov
\cite{Bykov,Vanstein,Vanbook} if we transform ellipsoidal to spheroidal coordinates, and have clear geometrical
interpretation. The integral for $S_1$ corresponds to the difference in lengths
of the two geodesic curves on $\eta_r$ between two points $P_1=(\xi_r,\eta_r,\phi_1)$ and
$P_2=(\xi_r,\eta_r,\phi_2)$. The first one goes from the caustic circle of intersection between
$\xi_r$ and $\eta_r$ along $\eta_r$ to the boundary surface $\xi_s$, reflects from it, and
returns back to the same circle. The second is simply the arc of the circle
between $P_1$ and $P_2$. The integral for $S_2$ corresponds to the length of a geodesic
line going from $P_1$ along $\xi_r$, lowering to $-\eta_r$ and returning to
$\eta_r$ at $P_2$ minus the length of the arc of the circle between $P_1$ and $P_2$. The
third integral is simply the length of the circle of intersection of $\xi_r$ and $\eta_r$.

These are elliptic integrals. For the whispering gallery modes when $\eta_r\ll 1$ and $\xi_0-\xi_r\ll \xi_r$, $S_2$
may be expanded into series over $\eta_r$ and $\zeta$ and integrated with the substitutions of $\eta=\eta_r\sin\psi$, $\zeta=(\xi^2-\xi_r^2)/\xi^2_r$ and Finally, expressing spheroidal coordinates $\xi_r$ and expressing $\xi_0$ through parameters of spheroid, we have:
\setlength{\arraycolsep}{0.0em}
\begin{eqnarray}
S_1&=&\frac{nb^3}{2a^2\sqrt{1+\zeta_0}}\int\limits \frac{\sqrt{\zeta}\sqrt{1+\zeta-\eta_r^2(1+\zeta_0)(b^2-a^2)/b^2}}{(1+\zeta_0+(\zeta-\zeta_0) b^2/a^2)\sqrt{1+\zeta}} d\zeta\nonumber\\
&=&\frac{nb^3}{2a^2\sqrt{1+\zeta_0}}\left[\frac{2}{3}\zeta^{3/2}-\frac{10a^2-4b^2}{15a^2}\zeta^{5/2}+\frac{a^2-b^2}{3b^2}\zeta^{3/2}\eta^2_r\right]\nonumber\\
&&+O(\zeta^{7/2},\eta_r^2\zeta^{5/2},\eta_r^4\zeta^{3/2})\nonumber\\
S_2&=&\frac{nd}{2}\eta^2_r\int\frac{\cos^2\psi\sqrt{\xi_r^2-s\eta_r^2\sin^2\psi}}{1-\eta_r^2\sin^2\psi}d\psi\\
&=&\frac{nd}{2}\eta^2_r\xi_r\left[\frac{2\psi+\sin 2\psi}{4}+ \frac{(2\xi_r^2-s)(4\psi-\sin4\psi)}{64\xi_r^2}\eta^2_r
\right.\nonumber\\
&+&\left.\frac{(8\xi_r^4-4f\xi_r^2-1)(12\psi+\sin6\psi-3\sin4\psi-3\sin2\psi)}{1536\xi_r^4}\eta^4_r\right.\nonumber\\
&+&\left.O(\eta^6_r) \right]\nonumber\\
S_3&=&\mu\phi,\nonumber
\end{eqnarray}
\setlength{\arraycolsep}{5pt}
Now we should solve the following system of equations:
\setlength{\arraycolsep}{0.0em}
\begin{eqnarray}
k\Delta S_1&=&2kS_1(\zeta_0)\nonumber\\
&\simeq&\frac{2\bar b^3nk\bar a}{3\bar a^3\sqrt{1+\zeta_0}} \zeta_0^{3/2}\left(1-\frac{5\bar a^2-2\bar b^2}{5\bar a^2}\zeta_0 -\frac{\bar b^2-\bar a^2}{2\bar b^2}\eta_r^2\right)\nonumber\\
&=&2\pi(q-1/4)\nonumber\\
k\Delta S_2&=&kS_2(2\pi)\nonumber\\
&\simeq&\pi \frac{nk\bar b}{\sqrt{1+\zeta_0}}\eta_r^2\left(1+\frac{\bar a^2+ \bar b^2}{8\bar b^2}\eta_r^2\right)\nonumber\\
&=&2\pi(p+1/2)\nonumber\\
k\Delta S_3&=&2\pi k\mu =2\pi\frac{nk\bar a}{\sqrt{1+\zeta_0}}\sqrt{1-\frac{\zeta_0(\bar b^2-\bar a^2)}{\bar a^2}}\sqrt{1-\eta^2_r}\nonumber\\
&=&2\pi|m|,
\end{eqnarray}
\setlength{\arraycolsep}{5pt}
Using the method of sequential iterations, starting for example from $nk^{(0)}a = l$, $\zeta_0^{(0)}=0$, $\eta_r^{(0)}=0$ this system may be resolved:
\setlength{\arraycolsep}{0.0em}
\begin{eqnarray}
\eta_r^2&=&\frac{(2p+1)a}{b}l^{-1}\left[1+\frac{\beta_q(b^2-a^2)}{2b^2}\left(\frac{l}{2}\right)^{-2/3}\right]+O(l^{-2})\nonumber\\
\zeta_0&=&\frac{-\beta_q a^2}{b^2}\left(\frac{l}{2}\right)^{-2/3}\left[1-\frac{\beta_q(5a^2-3b^2)}{5b^2}\left(\frac{l}{2}\right)^{-2/3}\right]\nonumber\\
&&+O(l^{-5/3})\nonumber\\
nk a &=& nk(\bar a-\sigma_r)=l-\beta_q\left(\frac{l}{2}\right)^{1/3}+\frac{2p(a-b)+a}{2b} \nonumber\\
&-&\frac{\chi n}{\sqrt{n^2-1}}+\frac{3\beta_q^2}{20}\left(\frac{l}{2}\right)^{-1/3}
\nonumber\\
&-&\frac{\beta_q}{12}\left(\frac{2p(a^3-b^3)+a^3}{b^3}+\frac{2n\chi(2\chi^2-3n^2)}{(n^2-1)^{3/2}}\right)\left(\frac{l}{2}\right)^{-2/3}\nonumber\\
&+&O(l^{-1}) \label{eikres},
\end{eqnarray}
\setlength{\arraycolsep}{5pt}
where for the convenience of comparison we introduced
$\beta_q=-[\frac{3}{2}\pi(q-\frac{1}{4})]^{2/3}$. The value of $\cos\theta$
needed for the calculation of $\Theta_r$ (\ref{Goos}) one may estimate as
$\cos\theta=\sqrt{1-l^2/(nk\bar a)}\simeq\beta l^{-1/3}$.

The first three terms for $nk a$ were obtained in \cite{microtorus,Bykov,Vanstein,Vanbook}
from different considerations, the last three are new.

To test this series we calculated using finite element method
eigenfrequencies of TE modes in spheroids with different eccentricities with totally
reflecting boundaries for $l=m=100$ (Fig.2). Significant improvement of our
series is evident. The divergence of the series for large eccenricities is
explained by the fact that the approximation that we used to calculate the
integrals (\ref{integrals}) but not the method itself breaks down in this case.
Namely $\eta_r$ becomes comparable to $\xi$ and should not be treated as a small
parameter.

\begin{figure}
\centering
\includegraphics[width=3.2in]{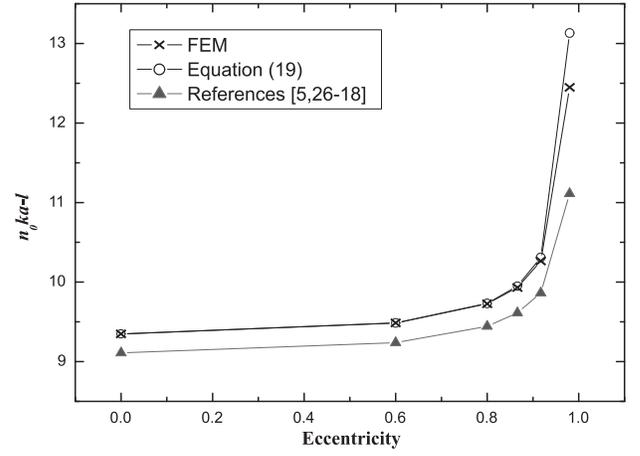}\\
\caption{Comparison of the precision of calculation of eigenfrequencies in spheroid.}
\end{figure}

If we put $a=b$, then all six terms in the obtained series coincide with that
obtained in \cite{Schiller} from exact solution in sphere with two minor
differences: 1) Airy function zeroes $\alpha_q\simeq  (-2.3381,-4.0879,-5.5206, ...)$ stand in
Schiller's series instead of approximate $\beta_q$ values ($\alpha_q-\beta_q\simeq$ -0.017; -0.0061; -0.0033, ...).
The reason is that the eikonal approximation breaks down on caustic, where
its more accurate extension with Airy functions is finite \cite{Keller}.
To make our solution even better we may just formally use $\alpha_q$ instead
of  $\beta_q$; 2) Minor difference in the last term is caused, we think, by
misprint in \cite{Schiller}, where in our designations instead of
$d_2=-2^{2/3}n^2 (-3+2\chi^2)\alpha_q/6$ should be $d_2=-2^{2/3}(-3n^2+2\chi^2)$.
The eikonal equation for the sphere may be solved explicitly and the expansion
of the solution shows that quasiclassical approximation breaks down on a term
$O(l^{-1})$, and of the same order should be the error introduced by substitution of vector equations by a scalar ones.

It is interesting to note that when $a=2b$ (oblate spheroid with the eccentricity
$\varepsilon =\sqrt{0.75}$, the eigenfrequency separation in the first order
of approximation between modes with the same $l$ determined by the third term
becomes equal to the separation between modes with different $l$ and the same
$l-m$ (free spectral range). The difference appears only in the term proportional
to $O(l^{-2/3})$. This situation is close to the case that was experimentally observed
in \cite{microtorus}. This new degeneracy has simple quasigeometrical interpretation
-- like in case of a sphere geodesic lines inclined to the equator plane on such
spheroid are closed curves returning at the same point of the equator after the whole
revolution, crossing, however, the equator not twice as big circles on a sphere but four times.

\section{Arbitrary convex body of revolution}

To find eigenfrequencies of whispering gallery modes in arbitrary body of
revolution one may use directly the results of the previous section by
fitting the shape of the body in convex equatorial area by equivalent
spheroid. In fact the body should be convex only in the vicinity of WG mode itself.
For example a torus with a circle of radius $r_0$ with its center at a distance $R_0$ from the $z$ axis
as a generatrix may be approximated by a spheroid with $a=R_0+r_0$ and $b=\sqrt{(R_0+r_0)r_0}$.
Nevertheless, more rigorous approach may be developed.

The first step is to find families of caustic surfaces. This is not a trivial
task in general but it is equivalent to finding caustic curves for the plane
curve forming the body of revolution which is in fact the so-called biinvolute
curve (the difference in length between a sum of two tangent lines from a point
on a curve to a biinvolute curve and an arc between these lines is constant).
Unfortunately we can not give now the formal provement of this statement but it looks true.
Another family form curves orthogonal to the first family. For example
in case of torus these families are concentric circles and radii,
that is why caustic surfaces for a torus are also concentric toruses.
In general case the following approximation may be used to find the first
family of biinvolute curves \cite{Babich}:
\begin{eqnarray}
n(s)\simeq\kappa\rho^{1/3}(s)+O(\kappa^2),
\end{eqnarray}
where $n(s)$ is the normal distance from a point $s$ on the curve to a biinvolute curve, $\kappa$ is a parameter of a family, and $\rho$ is the radius of curvature of the initial
curve at $s$.

Let we have found a caustic surface from the first family, parametrized as
\begin{eqnarray}
z&=&u \nonumber\\
x&=&g(u)\cos\phi \nonumber\\
y&=&g(u)\sin\phi
\end{eqnarray}

A geodesic line for this surface is given by the following integral:
\begin{eqnarray}
\frac{d\phi}{du}=c_1\frac{\sqrt{1+g'^2}}{g(u)\sqrt{g^2(u)-c_1^2}}
\end{eqnarray}
where $c_1$ is some constant, which is equal in our case to $\rho_r=g(z_{max})$ - the radius of caustic circle at maximum distance from equatorial plane.
The length of geodesic line:
\begin{eqnarray}
\frac{ds}{du}=\sqrt{1+g'^2+g^2\left(\frac{d\phi}{du}\right)^2}=\frac{g\sqrt{1+g'^2}}{\sqrt{g^2-\rho_r^2}}
\end{eqnarray}

The length of geodesic line, connecting points $\phi_1$ and $\phi_2$:
\begin{equation}
L^g_1= 2\int\limits_{-u_r}^{u_r}\frac{g\sqrt{1+g'^2}}{\sqrt{g^2-\rho_r^2}}du
\end{equation}
The length of arc from $\phi_0=0$ to $\phi_c=2\int_{-\eta_r}^{\eta_r}\frac{d\phi}{du}du$ is equal to $L^g_2=\rho_r\phi_c$.
\begin{equation}
L^g_2= 2\int\limits_{-u_r}^{u_r}\frac{\rho_r^2\sqrt{1+g'^2}}{g\sqrt{g^2-\rho_r^2}}du
\end{equation}
Finally:
\begin{eqnarray}
nk(L^g_1-L^g_2)&=& 2nk\int\limits_{-u_r}^{u_r}\frac{\sqrt{1+g'^2}\sqrt{g^2-\rho_r^2}}{g}du\nonumber\\
&=& 2\pi(p+1/2)\label{geod1}
\end{eqnarray}

In analogous way for another geodesic line on a caustic surface from the other family:
\begin{eqnarray}
z&=&v \nonumber\\
x&=&h(v)\cos\phi \nonumber\\
y&=&h(v)\sin\phi,
\end{eqnarray}
we have
\begin{eqnarray}
nk(L^h_1-L^h_2)&=& 2nk\int\limits_{v_r}^{v_0}\frac{\sqrt{1+h'^2}\sqrt{h^2-\rho_r^2}}{h}dv\nonumber\\
&=& 2\pi(q-1/4) \label{geod2}
\end{eqnarray}
The third condition is:
\begin{equation}
2\pi nk\rho_r= 2\pi|m|\label{geod3}
\end{equation}

With the substitution $u=\frac{d}{2}\xi\eta_r$, $v=\frac{d}{2}\xi_r\eta$, $g(u)=\frac{d}{2}\sqrt{\xi^2-s}\sqrt{1-\eta_r^2}$, $h(u)=\frac{d}{2}\sqrt{\xi_r^2-s}\sqrt{1-\eta^2}$ in
(\ref{geod1},\ref{geod2}, \ref{geod3}) we again obtain expressions for spheroid obtained before. For a torus
caustic surfaces are toruses and cones and are determined by the equations:
\begin{eqnarray}
g(u)&=& R_0+\sqrt{r_c^2-u^2} \nonumber \\
h(v)&=& R_0+ v\frac{\rho_r-R_0}{\sqrt{r_0^2-(\rho_r-R_0)^2}}
\end{eqnarray}

In conclusion. We have analyzed quasiclassical method of calculation of eigenfrequencies in spheroidal cavities and found that it gives approximations correct up to the term proportional to $l^{-2/3}$. This method may be easily expanded on arbitrary convex bodies of revolution.

\section*{Acknowledgment}
The work of M.L.Gorodetsky was supported in part by the Alexander von
Humboldt foundation.


\begin{thebibliography}{99}
\bibitem{microspheres} V. B. Braginsky, M. L. Gorodetsky and V. S. Ilchenko,
``Quality--factor and nonlinear properties of optical
whispering--gallery modes,'' {\it Phys.\ Lett.\ }{\bf A137,} pp. 393--397,
1989.
\bibitem{Richtmyer} R.D. Richtmyer,
``Dielectric Resonators'', {\it J. of Appl. Phys.} {\bf 10}, pp. 391--398, 1939.
\bibitem{microtorus} V.S.Ilchenko, M.L.Gorodetsky, X.S.Yao and L.Maleki,
``Microtorus: a high--finesse microcavity with whispering--gallery
modes'', {\it Opt.\ Lett.\ }{\bf 26}, pp. 256--258, 2001.
\bibitem{Vahala} K. Vahala,
``Optical microcavities'', {\it Nature\ }{\bf 424}, pp. 839--846, 2001.
\bibitem{crystall} V.S. Ilchenko, A.A. Savchenkov, A.B. Matsko et al.
``Nonlinear optics and crystalline whispering gallery mode cavities,'' {\it Phys.\ Rev.\ Lett.\ }{\bf 92,} (043903),
2004.
\bibitem{LEOS} {\it 2004 Digest of the LEOS Summer Topical Meetings: Biophotonics/Optical Interconnects \& VLSI Photonics/WGM Microcavities} (IEEE Cat. No.04TH8728), 2004.
\bibitem{Keller} J.B. Keller, S.I. Rubinow,
``Asymptotic solution of eigenvalue problems'', {\it Ann. Phys.} {\bf 9}, pp. 24--75, 1960.
\bibitem{Babich} V.M. Babi\u{c}, V.S. Buldyrev, {\it Short-wavelength
diffraction theory. Asymptotic methods}, Springer-Verlag, Berlin Heidelberg, 1991.
\bibitem{Schiller} S.Schiller, ``Asymptotic expansion of morphological resonance frequencies in Mie scatternig'', {\it Appl.\ Opt.\ },{\bf 32}, pp. 2181--2185, 1993.
\bibitem{TETM} R. Janaswamy,``A note on the TE/TM decomposition of electromagnetic fields in three dimensional homogeneous space'', {\it IEEE Trans. Antennas and Propagation}
{\bf 52}, pp. 2474--2477, 2004.
\bibitem{Asano} S.Asano, G.Yamamoto, ``Light scattering by a spheroidal particle'', {\it Appl.\ Opt.\ }{\bf 14}, pp. 29--49, 1975.
\bibitem{Farafonov} V.G. Farafonov, N.V. Voshchinnikov, ``Optical properties of spheroidal particles'', {\it Astrophys.\ and\ Space\ Sci.\ }{\bf 204}, pp. 19--86, 1993.
\bibitem{Charalambopoulos} A.Charalambopoulos, D.I.Fotiadis, C.V. Massalas, ``On the solution of boundary value problems using spheroidal eigenvectors'', {\it Comput.\ Phys. Comm.\ }{\bf 139}, pp. 153--171, 2001.
\bibitem{Moraes} P.C.G. de Moraes, L.G. Guimar\~{a}es,
``Semiclassical theory to optical resonant modes of a transparent dielectric spheroidal cavity'', {\it Appl.\ Opt.\ }{\bf 41}, pp. 2955--2961, 2002.
\bibitem{Closed} L. Li, Z. Li, M. Leong, ``Closed-form eigenfrequencies in prolate spheroidal conducting cavity'', {\it IEEE Trans. Microwave Theory Tech.} {\bf 51}, pp. 922--927, 2003.
\bibitem{Komarov} I. V. Komarov, L. I. Ponomarev, and S. J. Slavianov, {\it Spheroidal and Coulomb SpheroidalFunctions} (in russian) (Moscow:
), Nauka, Moscow, 1976.
\bibitem{Li}L.Li, X.Kang, M.Leong, {\it Spheroidal Wave Functions in Electromagnetic Theory}, John Wiley \& Sons, 2002.
\bibitem{Abramowitz} Handbook of Mathematical Functions, ed. M.Abramowitz and I.E.Stegun, National Bureau of Standards, 1964.
\bibitem{MoraesR} P.C.G. de Moraes, L.G. Guimar\~{a}es,
``Uniform asymptotic formulae for the spheroidal radial function'', {\it J. of Quantitative Spectroscopy and Radiative Transfer} {\bf 79--80}, pp. 973--981, 2003.
\bibitem{MoraesS} P.C.G. de Moraes, L.G. Guimar\~{a}es,
``Uniform asymptotic formulae for the spheroidal angular function'', {\it J. of Quantitative Spectroscopy and Radiative Transfer} {\bf 74}, pp. 757--765, 2003.
\bibitem{GDT} V.A.Borovikov, B.E.Kinber, {\it Geometrical theory of diffraction},
IEE Electromagnet. Waves Ser.37, London, 1994.
\bibitem{Silakov} E.L. Silakov,``On the application of ray method for the calculation of complex eigenvalues'' (in russian), {\it Zapiski nauchnogo seminara LOMI} {\bf 42}, pp. 228--235, 1974.
\bibitem{Born} M.Born and E.Wolf, {\it Principles of Optics}, 7-th ed., Cambridge University Press, 1999.
\bibitem{Arnold} V.I. Arnold, ``Modes and quasimodes'' (in russian), {\it Funktsionalny analiz i prilozheniya} {\bf 6}, pp. 12--20, 1972.
\bibitem{precess} M.L.Gorodetsky, V.S.Ilchenko,
``High-Q optical whispering-gallery microresonators: precession approach for spherical mode analysis and emission patterns with prism couplers'', {\it Opt.\ Commun.\ }{\bf 113}, pp. 133--143, 1994.
\bibitem{Bykov} V.P.Bykov,
``Geometrical optics of three-dimensional oscillations in open
resonators'' (in russian), {\it Elektronika bol'schikh moschnostei}
{\bf 4}, pp. 66--92, Moscow, 1965.
\bibitem{Vanstein} L.A. Vainstein,``Barellike open resonators'' (in russian), {\it Elektronika bol'schikh moschnostei} {\bf 3}, pp. 176--215, Moscow, 1964.
\bibitem{Vanbook} L.A. Vainstein, {\it Open Resonators and Open Waveguides}, Golem, Denver, 1969.
\end{thebibliography}
\end{document}